\documentclass[12pt]{iopart}

\usepackage{amsthm}
\usepackage{amssymb}
\usepackage{epsf}

\def\drawbox#1#2{\hrule height#2pt
        \hbox{\vrule width#2pt height#1pt \kern#1pt
              \vrule width#2pt}
              \hrule height#2pt}

\def\Asym#1#2{\vcenter{\vbox{\drawbox{#1}{#2}
              \kern-#2pt       % line up boxes
              \drawbox{#1}{#2}}}}

\def\href#1#2{#2}

\begin{document}

\thispagestyle{empty}

\rightline{CALT-68-2236}
\rightline{HUTP-99/A059}
\rightline{MIT-CTP-2891}
\rightline{hep-th/9910184}

\vspace{1.0truecm}
\centerline{\bf \Large Calabi-Yau Mirror Symmetry as a}
\vspace{.4truecm}
\centerline{\bf \Large 
Gauge Theory Duality 
\footnote[7]{to appear in the proceedings of Strings 99, Potsdam.}
}

\vspace{0.6truecm}
\centerline{\bf  Mina Aganagic
%\footnote{mina@theory.caltech.edu}
}
\vspace{.2truecm}
{\em \centerline{California Institute of Technology, Pasadena CA 91125}}
 {\em \centerline{and}}
{\em \centerline{Lyman Laboratory of Physics, Harvard University,
Cambridge, MA 02138}}
{\em\centerline{mina@cartan.harvard.edu}}
\vspace{.4truecm}
\centerline{\bf
Andreas Karch
%\footnote{karch@mit.edu}
}
\vspace{.2truecm}
{\em \centerline{Center for Theoretical Physics, MIT,
Cambridge, MA 02139}}
{\em\centerline{karch@ctp.mit.edu}}

\vspace{0.5truecm}

%%%%%%%%%%%%%%%%%%%%%%%%%%%%%%%%%%%%%%%%%%%%%%%%%%%%%%%%%

%\vspace{.4truecm}
\begin{abstract}
We show that there are two different dualities 
of two dimensional gauge theories with ${\cal N}=(2,2)$ supersymmetry. 
One is basically a consequence of 3d mirror symmetry.
The non-linear sigma model with Calabi-Yau target space on the Higgs
branch of the gauge theory is mapped into an equivalent 
non-linear sigma model on the Coulomb
branch of the dual theory, realizing a T-dual target space with torsion. The
second duality is genuine to two dimensions. In addition to swapping Higgs
and Coulomb branch it trades twisted for untwisted multiplets, implying a
sign flip of the left moving $U(1)_R$ charge. Succesive application 
of both dualities
leads to geometric mirror symmetry for the target space Calabi-Yau.
\end{abstract}

%\bigskip \bigskip

%%%%%%%%%%%%%%%%%%%%%%%%%%%%%%%%%%%%%%%%%%%%%%%%%%%%%%%%%%%%%

%       SECTION:  Introduction

%%%%%%%%%%%%%%%%%%%%%%%%%%%%%%%%%%%%%%%%%%%%%%%%%%%%%%%%%%%%%

\font\markfont=cmbxsl10 scaled \magstep1
\font\bigmarkfont=cmbxsl10 scaled \magstep3

\section{Introduction}
\setcounter{footnote}{1}
Mirror symmetry of Calabi-Yau manifolds is one of the remarkable
predictions of type II string theory. The $(2,2)$ superconformal
field theory associated with a string propagating on a Ricci-flat
K\"ahler manifold has a $U(1)_L\times U(1)_R$ R-symmetry
group, and the Hodge numbers of the manifold
correspond to the charges of 
$(R,R)$ ground states under the R-symmetry. There is a symmetry
in the conformal field theory which is a flip of the sign of
the $U(1)_L$ current, $J_L\rightarrow -J_L$. 
If physically realized, this symmetry implies existence
of pairs of manifolds $({\cal M},{\cal W})$, which 
have ``mirror'' hodge diamonds,  
$H^{p,q}({\cal M})=H^{d-p,q}({\cal W}),$ and give rise to exactly the
same superconformal field theory.

While this observation predicts existence of
mirror pairs of Calabi-Yau $d$-folds, it is not constructive:
one would like to know how to find ${\cal W}$ if one is given
a Calabi-Yau manifold ${\cal M}$.
The mirror construction has been proposed on purely mathematical
grounds by Batyrev for Calabi-Yau manifolds which can be 
realized as complete intersections in toric varieties.
 
In \cite{SYZ}, Strominger, Yau and Zaslow
argued that, for mirror symmetry to extend to a symmetry of
non-perturbative string theory, $({\cal M},{\cal W})$ must be
$T^{d}$ fibrations, with fibers which are special lagrangian,
and furthermore, that mirror symmetry is $T^{d}$-duality
of the fibers. The argument of SYZ is local, and is very well 
understood only for smooth fibrations.
To be able to fully exploit the idea, one must understand
degenerations of special lagrangian tori, 
or more generally, limits in which the Calabi-Yau manifold itself 
becomes singular.
Some progress on how this is supposed to work has been made in \cite{SYZ},
\cite{VafaLeung}, \cite{mina3}, and local mirror symmetry appears
to be the simplest to understand in this context.

In an a priori unrelated development initiated by \cite{mirror3d}
`mirror pairs' of gauge theories were found. In the field theory context, 
a duality in the sense of IR equivalence
of two gauge theories is usually referred to as a `mirror symmetry' if
\begin{itemize}
\item the duality swaps Coulomb and Higgs branch of the theories, trading
FI terms for mass parameters,
\item the R-symmetry of the gauge theory has the product form $G_L \times
G_R$ and duality swaps the two factors.
\end{itemize}
The example of \cite{mirror3d} studies ${\cal N}=4$ SUSY theories
in 3 dimensions. Recently it was shown \cite{KS} that this duality can 
even be generalized
to an equivalence of two theories at all scales,
a relation that will prove to be crucial for our applications.

The purpose of this note is to obtain geometrical mirror pairs by a
`worldsheet' construction. Since there seems
to be little hope that one can do so directly in the 
non-linear sigma model (NL$\sigma$M), we study
linear sigma models \cite{phases}
which flow in the IR to non-linear sigma models
with Calabi-Yau's as their target spaces. String theory
offers a direct physical interpretations of these models as
world-volume theory of a D-string probe of the Calabi-Yau manifolds.
We find gauge theory mirror duality for the linear sigma
model which reduces to geometric mirror symmmetry for Calabi-Yau target
spaces. 

Using brane constructions T-dual to the D1 brane probe on $\cal{M}$ one can
easily construct gauge theories which flow to mirror manifolds. 
We will show that 
two different dualities of $d=2$, ${\cal N}=(2,2)$
supersymmetric gauge theories arise:
One has a realization in string theory as `S-duality\footnote{
where by S-duality in type IIA setups we mean a flip of the 2 and the 10
direction}' of `interval' brane setups.
This duality is a consequence of mirror symmetry of $d=3$, ${\cal N}=2$ 
field theories
where both the Coulomb branch and the Higgs branch are described
by non-compact Calabi-Yau manifolds. This duality in $d=3$ maps a Calabi Yau
manifold to an identical one, so while it is a non-trivial field theory 
statement, this does
not provide a linear $\sigma$ model construction of the
mirror Calabi-Yau manifold. There exists another $d=2$, ${\cal N}=(2,2)$ field
theory duality obtained as S-duality of 
the diamond brane constructions of \cite{mina3}. This duality maps
a theory whose Coulomb branch is dual to Calabi-Yau manifold
${\cal M}$ in the ``boring'' way described above, to a theory whose Higgs
branch is the mirror manifold ${\cal W}$.
The composition of these
two dualities, therefore, flows to Calabi-Yau mirror symmetry.
While we consider in
this note a very particular family on non-compact Calabi-Yau manifolds,
the generalization to arbitrary affine toric varieties is 
possible.

The organization of the note is as follows:
In the next section we discuss different possible dualities in 
two dimensions as obtained via brane constructions for the case of the
conifold.
Section three
generalizes this discussion to sigma models built from branes dual to
more general non-compact CY manifolds and to the non-abelian case,
describing $N$ D-brane probes on the singular CY.
In the last section
we present a detailed study of the moduli space 
and argue that the composition of two dualities is Calabi-Yau mirror
symmetry.

\section{Mirror duality in 2d gauge theories}
\subsection{From three to two dimensions}
The original $D=3$ ${\cal N}=4$ of \cite{mirror3d} upon compactification 
implies also a duality relation in ${\cal N}=(4,4)$ theories in 2 dimensions, as noted
in \cite{sethim,brodiem}. The recent results of \cite{KS} 
are needed to make this
precise. The nature of this duality with 8 supercharges will 
teach us, how we should understand
the ${\cal N}=(2,2)$ examples. Since in 2d the concept of a moduli space is ill defined, 
equivalence of the IR physics does not require the moduli spaces and metrics to match point by
point, but only that the NL$\sigma$Ms on the moduli space\footnote{or in the non-compact
CY examples we are considering the two disjoint CFTs of Coulomb and Higgs branch \cite{higgsbranch,
comments}} are equivalent, as we will see in several examples.

Start with the 3d theory compactified on a circle. This is the setup analyzed in
\cite{diaconescu}. It is governed by two length scales, 
$g^{-2}_{YM}$, the 3d Yang-Mills coupling, and $R_2$, the compactification radius.
To flow to the deep IR is equivalent to sending both length scales to zero. However
physics still might depend on the dimensionless ratio
$$\gamma=g^2_{YM} \; R_2. $$
As shown in \cite{diaconescu}, while the Higgs branch metric is protected, the
Coulomb branch indeed does depend on $\gamma$. For $\gamma\gg 1$ we first have
to flow into the deep IR in 3d and then compactify, resulting in a 2d NL$\sigma$M
on the 3d quatum corrected Coulomb branch. The resulting target space is best described in terms of
the dual photon in 3d, a scalar of radius $\gamma$. For `the mirror of the quiver'
(U(1) with $N_f$ electrons) it turns out to be an ALF space with radius $\gamma$.
For small $\gamma$ we should first compactify, express the theory in terms of the Wilson line,
a scalar of radius $\frac{1}{\gamma}$, and obtain as a result a tube metric with torsion,
corresponding to the metric of an NS5 brane on a transverse circle
of radius $\frac{1}{\gamma}$ \cite{diaconescu,seibergsethi}.
Indeed these two NL$\sigma$Ms are believed to be equivalent \cite{NSdual} and exchanging
the dual photon for the Wilson line amounts to the T-duality of NS5 branes and ALF space
in terms of the IR NL$\sigma$M \footnote{This picture is obvious
from the string theory perspective. Studying a D2 D6 system on a circle, going to the IR first
lifts us to an M2 on an ALF space which becomes a fundamental string on the ALF, 
while going to 2d first makes us T-dualize to D1 D5, leaving us with the $\sigma$ model
of a string probing a 5-brane background.}. 
%Note that while the IR physics we end up with
%is always a 2d NL$\sigma$M model, the free UV theory (the linear $\sigma$ model) one has to use
%in order to obtain a CY (the ALF) on the Coulomb branch, is in fact 3d!
%This has to be contrasted with the situation on the Higgs branch, where $\gamma$ independence
%allows us to write down a purely 2d gauge theory in the UV \cite{phases}.

In order to obtain linear $\sigma$-model description of this scenario, one has to use the all scale mirror
symmetry of \cite{KS}. They show that $g^2_{YM}$ maps to a Fermi type coupling in the mirror
theory, or more precisely: one couples the gauge field via a BF coupling to a twisted gauge
field, the gauge coupling of the twisted gauge field
being baptised Fermi coupling. For the case
of the quiver theory with Fermi coupling, one obtains the same ALF space,
this time on the Higgs branch. 
In the same spirit we will present two
different dualities for ${\cal N}=(2,2)$ theories.

\subsection{Mirror symmetry from the interval}

One way to `derive' field theory duality is to embed
the field theory into string theory and then field theory duality
is a consequence of string
duality.
A construction of this sort was implemented in \cite{HW}
for the ${\cal N}=4$ theory in d=3  via brane configurations.
One uses an interval construction with
the 3 basic ingredients: NS5 along 012345, D5 along 012789
and D3 along 0126. The two R-symmetries are $SU(2)_{345}$ and
$SU(2)_{789}$. D3 brane segments between NS5 branes give rise
to vector multiplets, with the 3 scalars in the 3 of $SU(2)_{345}$.
D3 brane segments between D5 branes are hypermultiplets with the
four scalars transforming as 2 doublets of $SU(2)_{789}$. 

Under S-duality the D5 branes turn into NS5 branes and vice versa while D3
branes stay invariant. One obtains the same kind of setup but with
D5 and NS5 branes interchanged. S-duality of type IIB string theory is
mirror symmetry in the
gauge theory\footnote{
To be precise, the
S-dual theory
will really contain a gauge theory of twisted hypers coupled
to twisted vectors. 
If in addition one performs a rotation taking 345 into 789
space, the two R-symmetries are swapped and
the theory is written in terms of vectors and hypers.}.

Now let us move on to the 2d theories. 
The brane realization of this duality is via an interval theory in IIA
with NS and  NS' branes and D2 branes along 016 \cite{amihori}.
The IIA analog of S-duality, the 2-10 flip,
takes this into
D4 and D4' branes.
The following parameters define the interval brane setup
and the gauge theory:
\begin{itemize}
\item The seperation of NS and NS' brane along 7 is the FI term. It
receives a complex partner, the 10 seperation which maps to
the 2d theta angle.
\item The seperation of the D4 branes along 2 and 3 gives twisted masses to the
flavors.
\end{itemize}
Mirror symmetry maps the FI term to the twisted masses. A twisted mass sits in a background
vector multiplet and has to be contrasted with the standard mass from the superpotential which
sits in a background chiral multiplet. Like the real mass in ${\cal N}=2$
theories in $d=3$, it
arises from terms like
$$
 \int d^4 \theta Q^{\dagger} e^{V_B} Q.
$$
where $V_B$ is a background vector multiplet.

\noindent
{\bf An Example:}
As an example let us discuss the interval realization of the small resolution of the 
conifold. As shown in 
\cite{uranga,dandm} by performing
$T_6$ T-duality on a D-string probe of the conifold we get an 
interval realization of the conifold gauge theory in terms of an elliptic IIA setup
with D2 branes stretched on a circle with one NS and one NS' brane. 
In this IIA setup the seperation of the NS branes in 67 is
the small resolution, while turning on the diamond mode would be the deformation of the conifold.

The gauge group on the worldvolume of the D-string on the conifold is \cite{klebwit}
a $U(1) \times U(1)$ gauge group with 2 bifundamental flavors
$A_1$, $A_2$, $B_1$ and $B_2$.
We can factor out the decoupled center of mass motion, the diagonal
$U(1)$, which does not have any charged matter and hence is free.
We are left with an interacting $U(1)$ with 2 flavors.
The scalar in the decoupled vector multiplet is the position of the D1 brane in the
23 space transvere to the conifold.
While the Coulomb branch describes
seperation into fractional branes, the Higgs branch describes
motion on the internal space and reproduces the
conifold geometry. 
The complexified blowup mode
for resolving the conifold is the FI term and the $\theta$ angle.

After 2-10 flip, the dual brane setup is again 
an elliptic model, this time with 
one D4 and one D4' brane. 
The gauge theory is a single ${\cal N}=(8,8)$ $U(1)$
from the D2 brane with 2 additional ${\cal N}=(2,2)$ matter flavors 
from the D4 and D4' brane. That is we have
\begin{itemize}
\item 
3 `adjoints', that is singlet fields $X$, $Y$ and $Z$ and
\item  matter fields $Q$, $\tilde{Q}$, $T$ and $\tilde{T}$ with charges
+1,-1,+1,-1.
\item They couple via a superpotential
$$W=Q X \tilde{Q} + T X \tilde{T} .$$
\item The singlet $Z$ is decoupled and corresponds to the center of mass motion.
\end{itemize}
Turning on the FI term and the $\theta$ angle in the original theory is a 
motion of the NS brane along the 7 and 10 direction
respectively. It maps into a 23 motion for the D4 brane, giving a twisted mass to $Q$ and $\tilde{Q}$.
%The resulting mirror theory is summarized in
% table~\ref{t1}
%where
%it is also compared with the mirror which we obtain using diamonds.

This analysis can also be performed by going to the
T-dual picture of D1 branes probing D5 branes intersecting in codimension 2, that is
over 4 common directions.
Aspects of this setup and its T-dual cousins in various dimensions
have already been studied by numerous authors, e.g. for the
D3 D7 D7' system in \cite{sengp} or for the D0 D4 D4' in \cite{044}.
The resulting gauge theory agrees with what we have found by 
applying the standard interval rules.

\subsection{Twisted mirror symmetry from diamonds}

A second T-dual configuration for D1 brane probes of
singular CY manifolds is D3 branes ending on a curve of NS branes,
called diamonds in \cite{mina3}.
These setups are the $T_{48}$-duals of D1 brane probes of 
the ${\cal C}_{kl}$ spaces.
Indeed it was this relation that allowed us to derive the diamond matter
content to begin with \cite{mina3}.
In order to use the diamond construction to see mirror symmetry, we
use S-duality of string theory, as in the original work of \cite{HW}.
Let us first consider the
parameters defining a diamond and how they map under S-duality: 
\begin{itemize}
\item the complex parameter defining the NS brane diamond contains
the FI term which is paired up with
the 2d $\theta$ angle,
\item the S-dual D5-brane diamond is defined by a complex parameter
which is derived from a
{\it superpotential} mass term.
\end{itemize}
FI term and theta angle are contained in a
background twisted chiral multiplet.
Under the duality this twisted chiral multiplet is mapped to a
background chiral multiplet containing the mass
term. Since ordinary mirror symmetry mapped FI terms to mass terms in
twisted chiral multiplet,
the map of operators under
the two versions of duality will be different.

\noindent
{\bf An Example:}
Let us start once more with
the simplest example, the D1 string on the blowup of the conifold.
That is, we consider a single diamond, one NS and one NS' brane, on
a torus.
After S-duality
this
elliptic model with NS5 and NS5' brane turns
into an elliptic model with D5 and D5' brane.
Since we have only D-branes in this dual picture, the matter content can
be analyzed by perturbative string techniques.
To shortcut, we perform
T$_{48}$ duality to the D1 D5 D5' system
as in the interval setup.
For the special example of the conifold the two possible
mirrors do not differ in the gauge and matter content, only in the parameter
map. This will not be the case in the more general examples considered below.

As analyzed above,
the corresponding dual gauge theory is a U(1) gauge group with
3 neutral fields $X$, $Y$ and $Z$ and
two flavors $Q$, $\tilde{Q}$, $T$ and $\tilde{T}$
with charges $+1$, $-1$, $+1$, $-1$ respectively.
The superpotential in the singular case is
$ W=QX \tilde{Q} + TY \tilde{T}$.

By S-duality, as in the NS NS' setup,
turning on the D5-brane diamonds
corresponds turning on vevs for the d=4 hypermultiplets from
the D5 D5' strings.
Under ${\cal N}=(2,2)$ these hypermultiplets decompose into background chiral
multiplets and hence appear
as parameters in the superpotential.
If we call those chiral multiplets
$h$ and $\tilde{h}$ the corresponding superpotential
contributions are \cite{044} $Q h \tilde{T} + \tilde{Q} \tilde{h} T $,
so that all in all the full superpotential reads
$$ W=QX \tilde{Q} + TY \tilde{T}+ Q h \tilde{T} + \tilde{Q} \tilde{h} T.$$
%Both mirrors of the conifold,
%are summarized in table~\ref{t1}.

%\begin{table}[htb]
%\begin{tabular}{|c||c|c|c|}
%\hline
%&original theory& interval mirror&diamond mirror\\ 
%\hline
%\hline
%gauge group& $U(1)$&$U(1)$&$U(1)$ \\
%\hline
%matter content& 2 flavors &  2 flavors  & 
%2 flavors \\
%& $A_{1,2}$, $B_{1,2}$&$Q$, $T$, $\tilde{Q}$, $\tilde{T}$ & $Q$, $T$, $\tilde{Q}$, $\tilde{T}$ \\
%&&+ singlet fields $X$, $Y$ &+ singlet fields $X$, $Y$ \\
%\hline
%superpotential & $W=0$ & $W=X Q \tilde{Q} + Y T \tilde{T}$ & $W=X Q \tilde{Q} + Y T \tilde{T}$ \\
%%\hline
%smooth out & FI-term & twisted mass & $\Delta W=h Q \tilde{T} + \tilde{h} \tilde{Q} T$ \\
%singularity& and $\theta$ angle & for all flavors & \\
%\hline
%space time& Higgs branch & Coulomb branch & Coulomb branch \\
%part of&  Resolution of  &Resolution of  & Deformation of  \\
%moduli space& conifold&conifold&conifold\\
%\hline
%\end{tabular}
%\caption{The two mirror gauge theories
% of the linear sigma model on the conifold.}
%\label{t1}
%\end{table}

\section{More mirror pairs}

\subsection{Other singular CY spaces}
According to the analysis of \cite{uranga,mina3}, 
D1 brane probes on the blowup of spaces of the form
$$G_{kl}: \; \; xy=u^kv^l $$
are $T_6$ dual to an interval setup with $k$ NS and $l$ NS' branes. 
The gauge group is a $U(1)^{k+l-1}$ with
bifundamental matter. It is straight forward to construct 
interval mirrors via the 2-10 flip in terms
of a $U(1)$ with 2 singlets and $k+l$ flavors. The $k+l-1$ complexified FI terms map into the $k+l-1$ independend
twisted mass terms (one twisted mass can 
be absorbed by redefining the origin of the Coulomb branch).

Similarly we can construct diamond mirrors for D1 brane probes of 
$C_{kl}$ spaces,
$$
C_{kl}: \; \; xy=z^k, \; \; uv=z^l  .$$
The gauge group for the D1 brane probe is $U(1)^{2kl-1}$.
The mirror is once more a single $U(1)$ with 2 singlets
and $k+l$ flavors. 
This time $(k+1)(l+1)-3$ complexified FI terms map to superpotential masses.
Note that, while the D1-brane gauge theory has
$2kl-1$ FI terms, only 
$(k+1)(l+1)-3$ lead to independent deformations of the moduli space.
This is a consequence of the fact the 
D1 brane gauge theory is not the 
minimal linear sigma model of $C_{kl}$,
which is just a $U(1)^{(k+1)(l+1)-3}$
(the same phenomenon arises in the case
of $\mathbb{C}^3/\Gamma$ orbifolds \cite{dougmor}). 

\subsection{Generalization to non-abelian gauge groups}

Our realization in terms
of brane setups gives us for free the non-abelian version of the story, 
the mirror dual of $N$ D1 branes sitting
on top of the conifold.
Let us spell out the dual pairs once more in the simple example of the conifold. 
Generalization to arbitrary
$G_{kl}$ and $C_{kl}$ spaces is straight forward.
The gauge group on $N$ D1 branes on the blowup of the conifold is \cite{klebwit}
$$SU(N) \times SU(N) \times U(1) $$
where we already omitted the decoupled center of mass VM. The matter content consists of
2 bifundamental flavors $A_{1,2}$, $B_{1,2}$. They couple via a superpotential
$$W=A_1 B_1 A_2 B_2 - A_1 B_2 A_2 B_1.$$
The diamond 
mirror of this theory is a single $U(N)$ gauge groups with 3 adjoints
\footnote{And here by adjoint we really mean a $U(N)$ adjoint, that is a $SU(N)$ adjoint
and a singlet. The singlet in $Z$ once more corresponds to the overall center of mass
motion and decouples.} $X$, $Y$ and $Z$ and
2 fundamental flavors $Q$, $\tilde{Q}$, $T$, 
$\tilde{T}$ coupling via a superpotential:
$$W=X[Y,Z] + Q X \tilde{Q}+ T Y \tilde{T} + h Q \tilde{T} + \tilde{h} \tilde{Q} T$$
where $h$ and $\tilde{h}$ are the same background parameters determining the diamond as in the abelian case.

\section{Geometric mirror symmetry from linear sigma models}

The basic conjecture is that applying both dualities succesively maps the  
L$\sigma$M
for a given Calabi-Yau to the L$\sigma$M 
on the mirror.
The parameter map we have presented above
implies that the dual theory is formulated in terms of twisted multiplets, 
realizing the required flip in the R-charge.

In order to support our 
conjecture, 
let us do the calculation for the single D1 brane probe on a ${\cal C}_{kl}$
space.
By construction the Higgs branch of the gauge theory
we start with is
the blownup ${\cal C}_{kl}$ space.
The twisted mirror of this theory is a U(1) gauge theory coupled
to $k+l$ flavors $Q$, $\tilde{Q}$ and $T$, $\tilde{T}$ and two singlet
fields $X$ and $Y$.  The superpotential takes the form
\begin{equation}
\label{eq1}
W=\sum_{i=1}^k Q_i (X-a_i) \tilde{Q}^i + \sum_{a=1}^l T_a(Y-b_a) \tilde{T}^a
+ \sum_{ia} Q_i h^i_a \tilde{T}^a + \tilde{Q}^i \tilde{h}_i^a T_a,
\end{equation}
where $h$ and $\tilde{h}$ are background hypermultiplets parametrizing
the diamonds and the $a_i$ and $b_a$ are the relative 
positions of the D5 and D5' branes 
in the D1 D5 D5' picture along 45
and 89 respectively;
$\sum a_i = \sum b_a =0$. 

According to the conjecture we now must find the ordinary mirror
of this theory, whose Higgs branch, it is claimed, will be the
mirror manifold.
Ordinary mirror symmetry derives from 3d mirror symmetry. In three
dimensions the Higgs branch of the mirror theory is the same as
the quatum corrected Coulomb branch of the original one. For the
purpose of computing the mirror of the ${\cal C}_{kl}$ space
it suffices therefore to calculate the effective Coulomb branch
of the 3d U(1) gauge theory with $k+l$ flavors and superpotential
eq.(\ref{eq1}).

First let us study the classical moduli space.
The D-term equations require
$$\sum_{i=1}^k |Q_i|^2 - |\tilde{Q}^i|^2 +
\sum_{a=1}^l |T_a|^2 - |\tilde{T}^a|^2  =0$$
the F-term requirements for the $Q$, $T$, $\tilde{Q}$ and $\tilde{T}$ fields are
$$
N \left ( \begin{array}{c} Q\\T \end{array} \right ) =0, 
\;\;\;\;\;\;\;
(\tilde{Q}, \tilde{T}) N^T = 0
$$
where $N$ is the $k+l$ by $k+l$ matrix
\begin{equation*}
N= \left (
\begin{array} {c|c}
\mbox{diag} \{X-a_1, X-a_2, \ldots, X-a_k \} & h\\
\tilde{h} & \mbox{diag} \{Y-b_1, Y-b_2, \ldots, Y-b_l \}
\end{array}
\right )
\end{equation*}
In addition the scalar potential contains the standard piece
$$2 \sigma^2 (\sum_{i=1}^k |Q_i|^2 + |\tilde{Q}^i|^2 +
\sum_{a=1}^l |T_a|^2 + |\tilde{T}^a|^2) $$
from the coupling of the scalar $\sigma$ in the vector
multiplet to the matter fields and the F-terms for $X$ and $Y$. 
The classical Coulomb branch is three complex dimensional parametrized
by $X$, $Y$ and $\sigma + i \gamma$, where $\gamma$ is the dual
photon.
Along this branch,
$Q$, $\tilde{Q}$,
$T$ and $\tilde{T}$ are zero.
The  Coulomb branch meets the Higgs branch
along the curve
\footnote{Note that this is
the defining equation
of the curve the NS5-branes wrap, the diamond \cite{mina3}. It is 
also the defining equation of the complex structure of the local mirror manifold for
the blownup ${\cal C}_{kl}$, the deformed ${\cal G}_{kl}$, whose defining
equation obtained by adding the `quadratic pieces'
$UV - \mbox{det} (N)=0$
which do not change the complex structure.}
$$\mbox{det} (N)=0.
$$

Now consider the quantum Coulomb branch.
As shown in \cite{many} the quantum Coulomb branch of a 
$U(1)$ theory with $N_f=k+l$ flavors has an
effective description in terms of
chiral fields $V_+$ and $V_-$ and a superpotential
$$W_{eff}= - N_f (V_+ V_- \mbox{ det}(M))^{1/N_f}.$$
$M$ is the $k+l$ by $k+l$
meson matrix 
$$M =\left ( \begin{array}{cc} Q_i \tilde{Q}^j &
Q_i \tilde{T}^b \\ T_a \tilde{Q}_j & T_a \tilde{T}^b  
  \end{array} \right ) . $$
Far out on the Coulomb branch $V_{\pm}$ are related to the classical
variables via $V_{\pm} \sim e^{\pm 1/g^2 (\sigma + i \gamma)}$.
Adding the tree level superpotential
eq.(\ref{eq1}) written in the compact form $$\mbox{Tr } (N M)$$
to this effective superpotential,
the $M$ F-term equations describing our quantum Coulomb branch read
\begin{equation}
\label{modul}
\; \; \; \; \; \; \; \; \; \; \;
\; \; \; \; \; \; \; \; \; \; \;
N_{\beta \gamma} - (V_+ V_-)^{1/N_f} \frac{H_{\beta \gamma}}{\mbox{det}(M)^{
1-1/N_f}
}=0
\end{equation}
where $$H_{\beta \gamma} = \frac{\partial \mbox{det}(M)}{
\partial M^{\beta \gamma}}$$
Taking the determinant in eq.(\ref{modul}) we find
that the quantum Coulomb branch is described by a hypersurface
$$\mbox{det}(N)=V_+ V_-.$$
This is precisely the mirror manifold of ${\cal C}_{kl}$ \cite{mirror1}.
Since the origin $V_+=V_-=X=Y=0$
is no longer part of this branch of moduli space, we arrive
at a smooth solution even so we started from the effective superpotential
of \cite{many}
that is singular at the origin.

We here considered only mirror symmetry for ${\cal C}_{kl}$ spaces.
Since any affine toric CY can be imbedded in ${\cal C}_{kl}$ for 
sufficiently large $k$ and $l$, mirror symmetry for all
such spaces follows by deformation.

\section*{Acknowledgements}
We would like thank Jacques Distler, Ami Hanany, Sav Sethi,
Matt Strassler and Andy Strominger for
useful discussions.

\newpage

\bibliographystyle{utphys}
\bibliography{2dmirror}

\end{document}